%% file: BuildingModels_Basic_GreyBlackBox_PESGM2025_v2_ArXiv.tex
\documentclass[conference]{IEEEtran}
\IEEEoverridecommandlockouts
\usepackage{cite}
\usepackage{amsmath,amssymb,amsfonts}
\usepackage{algorithmic}
\usepackage{subcaption}
\usepackage{graphicx}
\usepackage{textcomp}
\usepackage{xcolor}
\def\BibTeX{{\rm B\kern-.05em{\sc i\kern-.025em b}\kern-.08em
    T\kern-.1667em\lower.7ex\hbox{E}\kern-.125emX}}

\usepackage{accents}
\newcommand{\ubar}[1]{\underaccent{\bar}{#1}}

\graphicspath{{./Images/}}

\begin{document}


\title{Comparing Building Thermal Dynamics Models and Estimation Methods for Grid-Edge Applications}

\author{
    Ninad Gaikwad\IEEEauthorrefmark{1}, Kunal Shankar\IEEEauthorrefmark{1}, Anamika Dubey\IEEEauthorrefmark{1}, Alan Love\IEEEauthorrefmark{2}, and Olvar Bergland\IEEEauthorrefmark{3}\\
    
    \IEEEauthorblockA{\IEEEauthorrefmark{1}School of Electrical Engineering \& Computer Science, Washington State University, Pullman, USA} 
    \IEEEauthorblockA{\IEEEauthorrefmark{2}School of Economic Sciences, Washington State University, Pullman, USA} 
    \IEEEauthorblockA{\IEEEauthorrefmark{3}School of Economics and Business, Norwegian University of Life Sciences, Ås, Norway} 
    \{ninad.gaikwad, kunal.shankar, anamika.dubey, a.love\}@wsu.edu, and olvar.bergland@nmbu.no
}

\maketitle


\begin{abstract}
We need computationally efficient and accurate building thermal dynamics models for utilization in grid-edge applications. This work evaluates two grey-box approaches for modeling building thermal dynamics: RC-network models and structured regression models. For RC-network models, we compare parameter estimation methods—Nonlinear Least Squares, Batch Estimation, and Maximum Likelihood Estimation. We utilize Almon Lag Structure with Linear Least Squares for the estimation of the structured regression models. We evaluate the performance of these models and methods on simulated house and commercial building data for three different simulation types.
\end{abstract}




\section{Introduction}\label{sec:Introduction}
According to the US Energy Information Administration (EIA), buildings account for $40\%$ of the total electricity consumed during the year 2022; out of which approximately $45\%$ is used for Heating Ventilation and Air-Conditioning (HVAC), see~\cite{EIAEnergyConsumptionWebpage}. This substantial HVAC energy usage can be tapped as a grid-edge resource by leveraging the thermal inertia of buildings and modulating HVAC operations to provide ancillary grid support. Developing and testing such grid-edge applications require computationally efficient and accurate thermal dynamic models. While, white-box models offer detailed physics-based representations but are computationally prohibitive for large-scale simulations, and black-box models lack physical interpretability; grey-box models combine physics-based formulations with data-driven estimation, providing a balance between computational efficiency and interpretability~\cite{li2021grey}.

One of the common approaches for grey-box modeling is through RC-network models, which represent building thermal dynamics using equivalent electrical circuits based on the fundamental physics principle of heat flow from higher to lower temperatures \cite{yang2024review}. Estimating the parameters of these models is challenging as it involves solving nonlinear optimization problems, which can be formulated in various ways. Nonlinear Least Squares (NLS) formulation has been applied to commercial buildings in~\cite{dewson1993least} and to residential buildings in~\cite{cui2019hybrid}. Batch Estimation (BE) formulation provided in~\cite{robertson1996moving}, has been explored for both residential and commercial buildings in~\cite{kuhl2011real} and~\cite{guo2019identification} respectively. A third formulation based on Maximum Likelihood Estimation (MLE) provided in~\cite{simpson2023efficient}, to the best of our knowledge, has not been directly applied to building thermal dynamics. On the other hand, regression models provide another way to model building thermal dynamics; these models usually follow an autoregressive form with exogenous inputs, and estimation is done using Linear Least Squares (LLS). Regression-based models have been applied to residential buildings in~\cite{hure2022regression} and commercial buildings in ~\cite{wu2012multi}, demonstrating their effectiveness in capturing key thermal behaviors in a simpler and computationally efficient manner.

This work makes the following contributions: (i) proposes a framework for modeling building thermal dynamics for grid-edge applications, (ii) compares parameter estimation techniques for RC-network models— NLS, BE, and MLE, (iii) introduces the Almon Lag Structure (ALS) for estimating regression based building thermal dynamics models, (iv) evaluates and compares the performance of these estimation methods on both residential and commercial buildings.


\section{Building Model for Grid-Edge}\label{sec:Modeling}
A general building model for grid-edge applications is described as follows;
\begin{align}
    \ubar x(k+1) &=  f_{Build}(\ubar x(k), \ubar u(k), \ubar w(k) ; \ubar \theta), \label{eq:GM1} \\
    Q_{HVAC}(k) &=  f_{Q_{HVAC}}(\ubar x(k), \ubar u(k)), \label{eq:GM2} \\ 
    P_{HVAC}(k) &= f_{P_{HVAC}}(|Q_{HVAC}(k)|), \label{eq:GM3} \\
    P(k) &=   P_{HVAC}(k) +   P_{Other}(k), \label{eq:GM4} \\
    Q(k) &=  P(k)tan(cos^{-1}(p.f(k))). \label{eq:GM5}
\end{align}
The general building model for grid-edge applications is structured as equations that describe building thermal states, HVAC energy use, and power consumption where $t$ is the discrete time-step index. The state evolution eq.~(\ref{eq:GM1}) defines the next state $\ubar x(k+1)$ through a dynamics function $f_{Building}$, based on the current state $\ubar x(k)$, control inputs $\ubar u(k)$, disturbances $\ubar w(k)$, and parameters $\ubar \theta$. The HVAC heat transfer rate, $Q_{HVAC_{i}}(k)$, is given in eq.~(\ref{eq:GM2}) as a function of thermal state and control inputs, representing the thermal load required for indoor climate control, the difference between commercial and residential buildings arise in the modeling of this equation as it depends on the specific type of HVAC system which are different for commercial and residential buildings. The HVAC power, $P_{HVAC}(k)$, is computed from $|Q_{HVAC}(k)|$ (absolute value is required as HVAC heat transfer rate can assume both positive and negative values depending on heating and cooling operation) in eq.~(\ref{eq:GM3}), while total power consumption $P(k)$ is the sum of HVAC and other power loads in eq.~(\ref{eq:GM4}). Finally, reactive power $Q(k)$ is derived from $P(k)$ and the building’s power factor $p.f(k)$ in eq.~(\ref{eq:GM5}). Together, these equations offer a framework for modeling buildings in grid-edge applications.

\section{Building Thermal Dynamics Modeling}\label{sec:Modeling}

\subsection{Thermal RC-Network Model}\label{subsec:RCModeling}
Thermal RC-network models are predominantly utilized to describe building thermal dynamics and are based on the physics principle that heat flows from higher-temperature to lower-temperature regions. These models take the form of electrical circuits where nodes represent temperatures (analogous to voltage), the resistances represent the thermal resistance that is offered to heat flow (analogous to current) between two nodes, and the capacitances represent the thermal capacitance that does not allow nodes to instantaneously change their temperature. A general RC-network model for an aggregated building thermal dynamics model is given as:
\begin{align}
\small
    &C_z \frac{dT_z}{dt} = \sum_{i=1}^{N} \frac{T_{w_{i}} - T_{z}}{R_{zw_{i}}} + \frac{T_{am} - T_{z}}{R_{za}} + A_z Q_{HVAC} \label{eq:GB1} \\ 
    &\quad \quad \quad + B_z Q_{Int} + D_z Q_{Solar}, \nonumber \\
    & C_{w_{i}} \frac{dT_{w_{i}}}{dt} = \sum_{\substack{j=1 \\ j \neq i}}^{N} \frac{T_{w_{j}} - T_{w_{i}}}{R_{w_{ij}}} + \frac{T_z - T_{w_{i}}}{R_{zw_{i}}} + \frac{T_{am} - T_{w_{i}}}{R_{wa_{i}}} \label{eq:GB2} \\
    &\quad \quad \quad + B_{w_{i}} Q_{Int} + D_{w_{i}} Q_{Solar}. \nonumber 
\end{align}
where, $T_{z}$, $T_{w_{i}}$, and $T_{am}$ are the aggregated indoor air temperature of a building, the temperature of the $i^{\text{th}}$ unmeasurable state out of $N$ (these are usually temperatures of fictitious surfaces that enclose the mathematically aggregated building model), and the ambient outside temperature respectively. $Q_{HVAC}$, $Q_{Int}$ and $Q_{Solar}$ are the heat accepted (cooling operation) or rejected (heating operation) by the HVAC system, the heat gain from solar irradiance, and the heat gain from internal heat sources (people, electrical equipment, etc.) respectively. $A_z$, $B_z$-$B_{w_{i}}$, $D_z$-$D_{w_{i}}$ are the proportions of $Q_{HVAC}$, $Q_{Int}$ and $Q_{Solar}$ affecting the $T_{z}$ and $T_{w_{i}}$ respectively.

We can write the eq.s (\ref{eq:GB1}) and (\ref{eq:GB2}) as a general linear state-space model given as;
\begin{small}
\begin{align}
    \dot{\ubar{x}} &=  A( \ubar \theta) \ubar{x} + B(\ubar \theta) \ubar{u} + D(\ubar \theta) \ubar{w} \label{eq:LSS1} \\
    y &= C \ubar{x} \label{eq:LSS2}.
\end{align}
\end{small}
where, the state is $\ubar{x} = [T_{z}, T_{w_{1}}, \dots, T_{w_{N}}]^{T}$, the control is $\ubar{u} = Q_{HVAC}$, the disturbances are $\ubar{w} = [T_{am}, Q_{Int}, Q_{Solar}]^{T}$, and the output $y = T_{z}$. The $A(\ubar \theta)$, $B(\ubar \theta)$, and $D(\ubar \theta)$ are the system matrix, input matrix, and disturbance matrix respectively; where $\ubar \theta = [R_{za}, R_{zw_{i}}, \dots, R_{zw_{N}}, R_{w_{1,2}}, \dots, R_{w_{N-1, N}}, R_{wa_{i}}, \dots, R_{wa_{N}},$ $ C_{z}, C_{w_{i}}, \dots, C_{w_{N}}, A_{z}, B_{z}, B_{w_{1}}, \dots, B_{w_{N}}, D_{z}, D_{w_{1}}, \dots,$ $D_{w_{N}}  ]^{T}$ is the parameter vector which we need to learn from data obtained from the system. $C = [1, 0, \dots, 0]$ is the output matrix as the only measurable state of the system is $T_{z}$. It should be noted that the model complexity and expressivity increase with $N$, and the eq.~(\ref{eq:GB1}) under an appropriate time discretization scheme become the $f_{Build}(\ubar x(k), \ubar u(k), \ubar w(k) ; \ubar \theta)$.

\subsection{Structured Regression Model}\label{subsec:RegModeling}
Here, we look at a regression-based model to represent the thermal dynamics of buildings based on a structured autoregressive model with exogenous inputs described in~\cite{manandhar2022dynamic} and~\cite{manandhar2022autonomous}. The general form of this model is given as;
\begin{small}
\begin{align}
    &T_{z}(k+1) = \alpha_0 + \sum_{i=l}^{t} \beta_i  T_{z}(k-i) + \sum_{i=j}^{m} \gamma_i  P_{c}(k-i)   \label{eq:StructuredReg_1} \\ &+ \sum_{i=j}^{m} \delta_i  P_{h}(k-i)  
     + \sum_{i=n}^{o} \eta_i  D_{c}(k-i) + \sum_{i=n}^{o} \vartheta_i  D_{h}(k-i) . \nonumber
\end{align}
\end{small}
Where, $P_{c}$ and $P_{h}$ are the HVAC power usage during cooling and heating operation respectively. $D_{c}  \triangleq \max(0, T_{am} - 19.44^{\circ}C)$ and $D_{h} \triangleq \max(0, 19.44^{\circ}C - T_{am})$ are the outside cooling and heating degrees respectively. $l, j, n $ and $t, m, o$ are the starting and ending lags respectively. Now, if we define $\ubar x \triangleq [T_{z}(k-l), \dots, T_{z}(k-t)]^{T}$, $\ubar u \triangleq [P_{c}(k-j), \dots, P_{c}(k-m), \dots, P_{h}(k-j), \dots, P_{h}(k-m)]^{T}$, $\ubar w \triangleq [D_{c}(k-n), \dots, D_{c}(k-o), \dots, D_{h}(k-n), \dots, D_{h}(k-o)]^{T}$, $y \triangleq T_{z}(k)$, and we have $\ubar \theta = [\alpha_{0}, \beta_{0}, \dots, \beta_{t-l},\gamma_{0}, \dots, \gamma_{m-j},\delta_{0}, \dots, \delta_{m-j},\eta_{0}, \dots, \eta_{o-n}, $ $ \vartheta_{0}, \dots, \vartheta_{o-n},]^{T}$; then eq.~(\ref{eq:StructuredReg_1}) represents the $f_{Build}(\ubar x(k), \ubar u(k), \ubar w(k) ; \ubar \theta)$. It should be noted that both the control input (HVAC power usage) and disturbance (ambient temperature) in this model are easily measurable with sensors as opposed to the thermal RC-network model.


\section{Model Estimation Methods}\label{sec:Methods}
Our goal is to estimate the parameters ($\ubar{\theta}$) of the models described in Section~\ref{sec:Modeling}. We use $T$ samples of the input-output data tuples collected in a dataset $\mathbb{D} = \{ (\ubar{u}(1), \ubar{w}(1), y(1)), \dots, (\ubar{u}(T), \ubar{w}(T), y(T)) \}$, where, $\ubar{u}$, $\ubar{w}$ are measured inputs, and $y$ is measured output of the system.

\subsection{Estimation Methods for Thermal RC-Network Model}\label{subsec:RCEstimationMethods}
The thermal RC-network model given in ~(\ref{eq:GB1}) and (\ref{eq:GB2}) is nonlinear w.r.t. the parameter vector $\ubar{\theta}$. We utilize three commonly used methods to estimate $\ubar{\theta}$ while using Euler discretization of the continuous-time dynamics.

\subsubsection{Nonlinear Least Squares}\label{subsubsec:NLSMethod}
NLS is described in (\ref{eq:LS_objective}) - (\ref{eq:LS_output}). The objective is to minimize the square of the difference between the actual output measurement and the predicted output, while constrained by the thermal RC-network dynamics.
\begin{small}
\begin{align}
    & \min_{\ubar x, \ubar \theta} \quad \sum_{k=1}^{T} \left( y(k) - \tilde{y}(k) \right)^2 \label{eq:LS_objective} \\
    &\text{subject to:}  \nonumber \\
    & \ubar{x}(k+1) = \ubar{x}(k) + t_s [ A( \ubar \theta) \ubar{x}(k) + B(\ubar \theta) \ubar{u}(k)  + D(\ubar \theta) \ubar{w}(k) ],   \nonumber \\
    & \tilde{y}(k) = C \ubar{x}(k) .  \label{eq:LS_output}
\end{align}
\end{small}

\subsubsection{Batch Estimation}\label{subsubsec:BEMethod}
BE formulation is given in (\ref{eq:BE_objective}) -(\ref{eq:BE_residual}). It incorporates the effects of process noise and measurement noise in the estimation model, see~\cite{robertson1996moving}.
\begin{align}
\small
    & \min_{\ubar w_{n}, v_{n}, \ubar x, \ubar \theta} \quad (\ubar x^{e}_{0})^T P_0^{-1} \ubar x^{e}_{0} + \sum_{k=0}^{T} \left( v_{n}(k)^T R^{-1} v_{n}(k) \right) \label{eq:BE_objective} \\ 
    & \quad \quad \quad + \sum_{k=0}^{T-1} \left( \ubar w_{n}(k)^T Q^{-1} \ubar w_{n}(k) \right) \nonumber \\
    & \text{subject to:}  \nonumber \\    
    & \ubar x(k+1) = \ubar{x}(k) + t_s [A( \ubar \theta) \ubar{x}(k) + B(\ubar \theta) \ubar{u}(k) \label{eq:BE_state_dynamics} \\
    & \quad \quad \quad \quad + D(\ubar \theta) \ubar{w}(k) ] + \ubar w_{n}(k), \nonumber \\
    & v_{n}(k) = y(k) - C \ubar{x}(k) . \label{eq:BE_residual} 
\end{align}
where, $\ubar x^{e}_{0} \triangleq \ubar x (0) - \ubar x_{0}$ is the initial state error estimate, here $\ubar x_{0}$ is the initial state estimate, and $P_0$ is the estimate of the initial state error covariance matrix. $\ubar w_{n} \sim N(\ubar{0}, Q)$ is the process noise with covariance matrix $Q$, and $ v_{n} \sim N(0, R)$ is the measurement noise with variance $R$. During the implementation $Q$ and $R$ are utilized as hyperparameters.

\subsubsection{Maximum Likelihood Estimation}\label{subsubsec:MLEMethod}
MLE formulation also incorporates the effect of process and measurement noise; however, it does it by introducing the Kalman-Filter version of the thermal RC-network dynamics in the constraints and minimizing the one-step prediction error ($e$) computed in the filtering step. The MLE formulation is described in (\ref{eq:MLE_objective}) -(\ref{eq:MLE_S}), see~\cite{simpson2023efficient}. 
\begin{small}
\begin{align}
    & \min_{P, S, \ubar e, \ubar{\tilde{x}}, \ubar \theta} \quad 
   \sum_{k=1}^{T} \ubar e(k)^{T} S(k)^{-1} \ubar e(k) + \log \mid S(k) \mid \label{eq:MLE_objective}\\
    & \text{subject to:}  \nonumber \\
    & \ubar{\tilde{x}}(k+1) = (I + t_{s} I) A( \theta ) [ \ubar{ \tilde{x} }(k) + P(k) C^T S(k)^{-1}  ) \label{eq:MLE_state_update}\\
    & \quad \quad  \quad \quad  \quad e(k) ] + t_{s} B(\theta) \ubar{u(k)} +   D(\theta) \ubar{w(k)} ], \nonumber \\
    & C(k+1) = (I + t_{s} I) A(\theta) [ P(k) - P(k) C^T S(k)^{-1}  \label{eq:MLE_covariance_update} \\
    & \quad \quad  \quad \quad  \quad  C P(k)] (I + t_{s} I) A(\theta)^T + Q, \nonumber  \\
    & \ubar e(k) = y(k) - C \ubar{\tilde{x}}(k), \label{eq:MLE_residual} \\
    & S(k) = C P(k) C^T + R \label{eq:MLE_S}.
\end{align}
\end{small}
where, $\ubar{ \tilde{x} }$, $P$ and $S$ are the predicted state, state error covariance matrix, and the one-step error variance respectively.

\subsection{Estimation Method for Structured Regression Model}\label{subsec:StructuredRegEstimationMethod}
The model in eq.~(\ref{eq:StructuredReg_1}) is linear in the parameters ($\ubar{\theta}$); hence, the estimation can be performed through the Linear Least Squares (LLS) method. However, directly using LLS on eq.~(\ref{eq:StructuredReg_1}) is problematic; as the inclusion of multiple high-frequency lags of the same regressor leads to problems with multicollinearity causing imprecise parameter estimates. To address this problem,~\cite{almon1965distributed} proposed a solution of applying polynomial restrictions (Almon Lag Structure - ALS) for the lagged weights (not including $\alpha_{0}$) reducing the parameter search space. For an arbitrary lagged regressor $z(k-i)$ for $i \in \{ l,\dots,t \}$ its associated lagged weight $\zeta_{i}$ can be restricted by using $q+1$ ALS parameters $(\omega_{\zeta,0},\dots,\omega_{\zeta,q})$ through the polynomial function given as $\zeta_{i} = \sum_{j=0}^{q} \omega_{\zeta,j} \, i^{j}$. The transformation of $z$ through the ALS method can be represented in a compact manner as $\text{A}(z,l,t,q)$ and is given in eq.~(\ref{eq:ALS}), and the ALS transformation of eq.~(\ref{eq:StructuredReg_1}) is given in eq.~(\ref{eq:ALS_Transform}). 
\begin{align}  
    & \text{A}(z,l,t,q) = \sum_{i = l}^{t}  \sum_{j=0}^{q} \omega_{\zeta,j} \; z(k-i)  , \label{eq:ALS} \\
    &T_{z}(k+1) = \alpha_0 + \text{A}(T_{z},l,t,q_{t}) + \text{A}(P_{c},j,m,q_{p})    \label{eq:ALS_Transform} \\ &+ \text{A}(P_{h},j,m,q_{p})  
     + \text{A}(D_{c},n,o,q_{d}) + \text{A}(D_{h},n,o,q_{d}) . \nonumber 
\end{align}
Note that (\ref{eq:ALS_Transform}) is linear in the ALS parameters, hence their estimation can proceed through LLS. It is important to note that maintaining $q < t-l+1$ leads to lesser parameters to be estimated for the ALS transformed equation, and $q$ acts as a hyperparameter.
 

\begin{figure*}[h]
	\centering
	\begin{subfigure}[t]{0.16\textwidth}
		\centering
		\includegraphics[width=\textwidth]{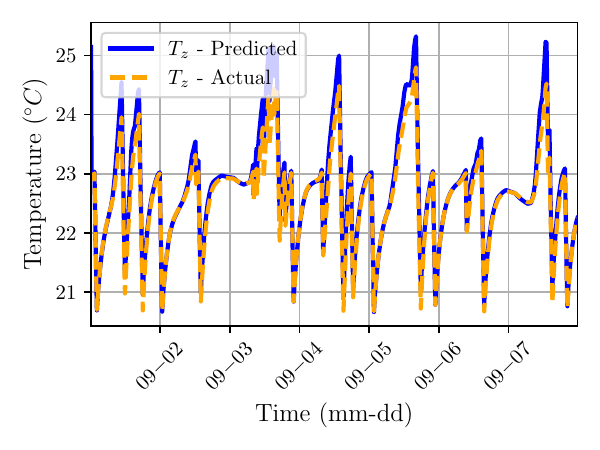}
		\caption{LS Sim1 R-4.}
		\label{fig:LS_Sim1_RES4}
	\end{subfigure}
	\begin{subfigure}[t]{0.16\textwidth}
		\centering
		\includegraphics[width=\textwidth]{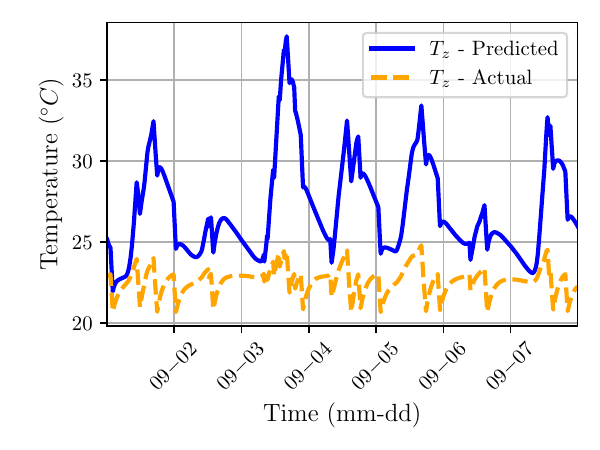}
		\caption{LS Sim2 R-4.}
		\label{fig:LS_Sim2_RES4}
	\end{subfigure}
	\begin{subfigure}[t]{0.16\textwidth}
		\centering
		\includegraphics[width=\textwidth]{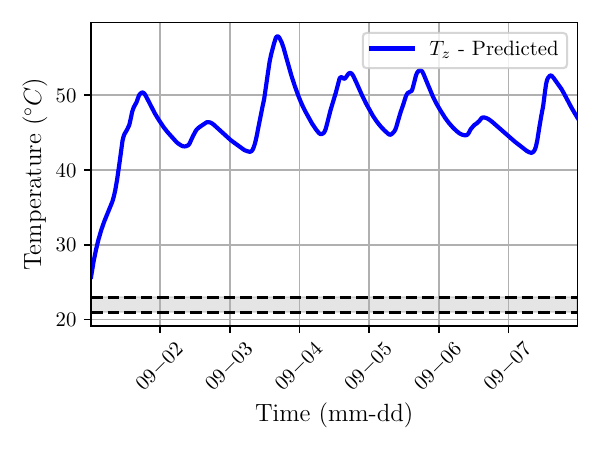}
		\caption{LS Sim3 R-4.}
		\label{fig:LS_Sim3_RES4}
	\end{subfigure}
	\begin{subfigure}[t]{0.16\textwidth}
		\centering
		\includegraphics[width=\textwidth]{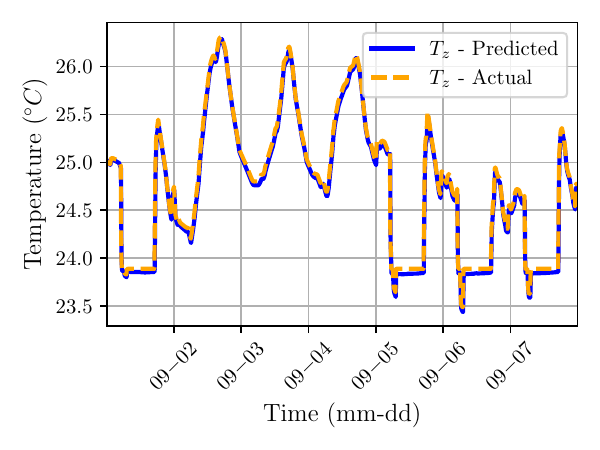}
		\caption{LS Sim1 C-4.}
		\label{fig:LS_Sim1_COM4}
	\end{subfigure}
	\begin{subfigure}[t]{0.16\textwidth}
		\centering
		\includegraphics[width=\textwidth]{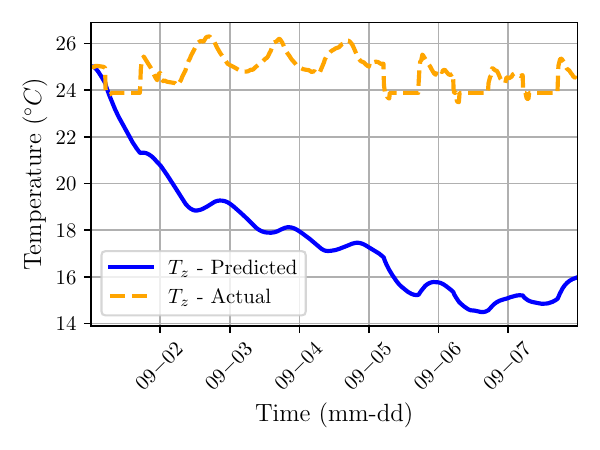}
		\caption{LS Sim2 C-4.}
		\label{fig:LS_Sim2_COM4}
	\end{subfigure}
	\begin{subfigure}[t]{0.16\textwidth}
		\centering
		\includegraphics[width=\textwidth]{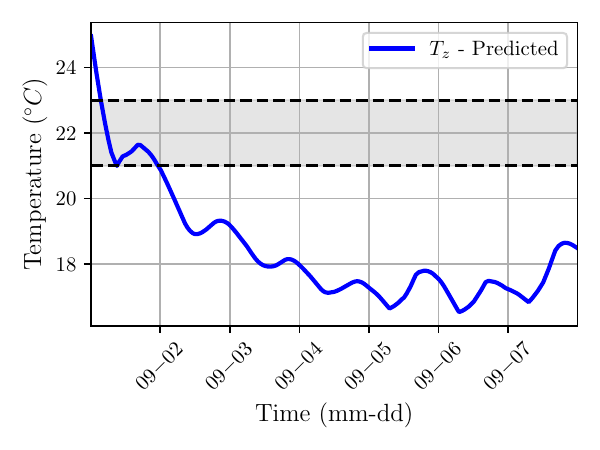}
		\caption{LS Sim3 C-4.}
		\label{fig:LS_Sim3_COM4}
	\end{subfigure}

	\begin{subfigure}[t]{0.16\textwidth}
		\centering
		\includegraphics[width=\textwidth]{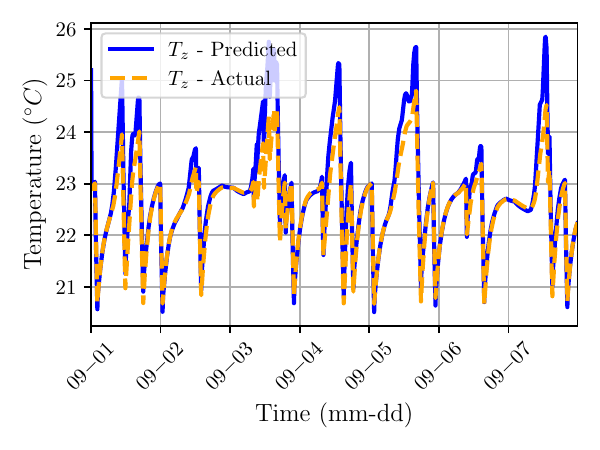}
		\caption{BE Sim1 R-4.}
		\label{fig:BE_Sim1_RES4}
	\end{subfigure}
	\begin{subfigure}[t]{0.16\textwidth}
		\centering
		\includegraphics[width=\textwidth]{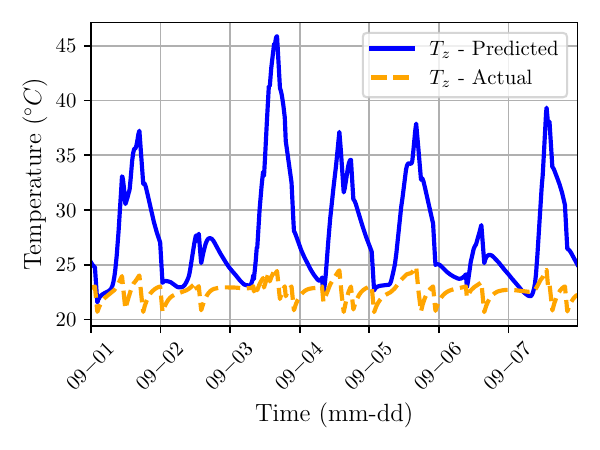}
		\caption{BE Sim2 R-4.}
		\label{fig:BE_Sim2_RES4}
	\end{subfigure}
	\begin{subfigure}[t]{0.16\textwidth}
		\centering
		\includegraphics[width=\textwidth]{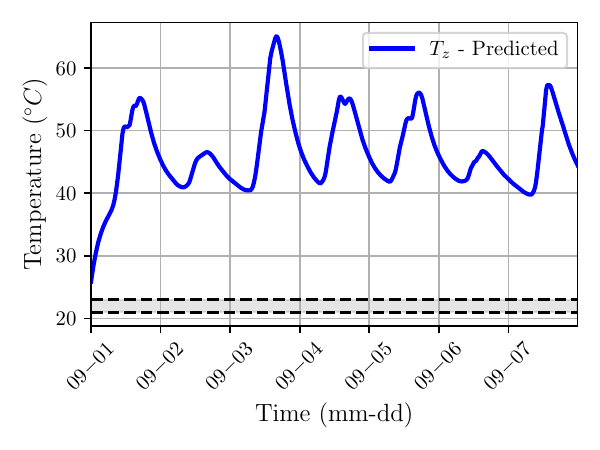}
		\caption{BE Sim3 R-4.}
		\label{fig:BE_Sim3_RES4}
	\end{subfigure}
	\begin{subfigure}[t]{0.16\textwidth}
		\centering
		\includegraphics[width=\textwidth]{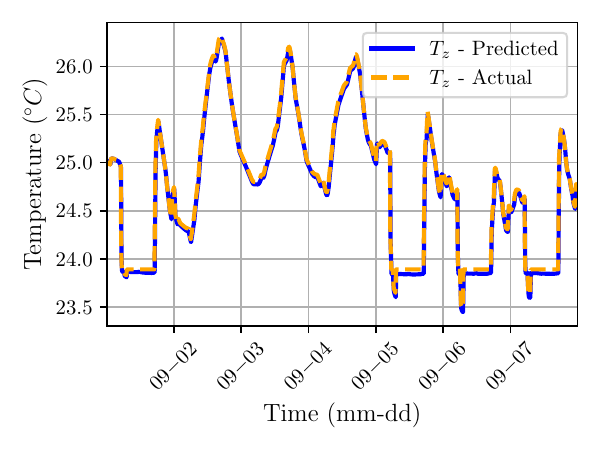}
		\caption{BE Sim1 C-4.}
		\label{fig:BE_Sim1_COM4}
	\end{subfigure}
	\begin{subfigure}[t]{0.16\textwidth}
		\centering
		\includegraphics[width=\textwidth]{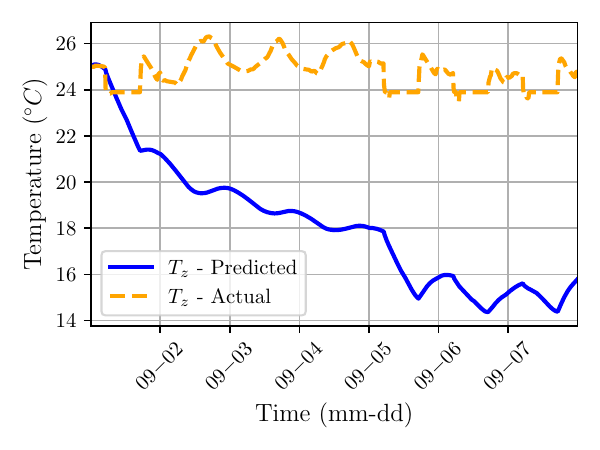}
		\caption{BE Sim2 C-4.}
		\label{fig:BE_Sim2_COM4}
	\end{subfigure}
	\begin{subfigure}[t]{0.16\textwidth}
		\centering
		\includegraphics[width=\textwidth]{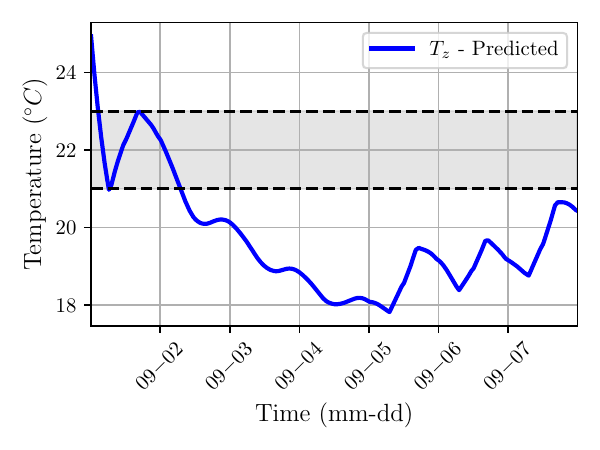}
		\caption{BE Sim3 C-4.}
		\label{fig:BE_Sim3_COM4}
	\end{subfigure}

	\begin{subfigure}[t]{0.16\textwidth}
		\centering
		\includegraphics[width=\textwidth]{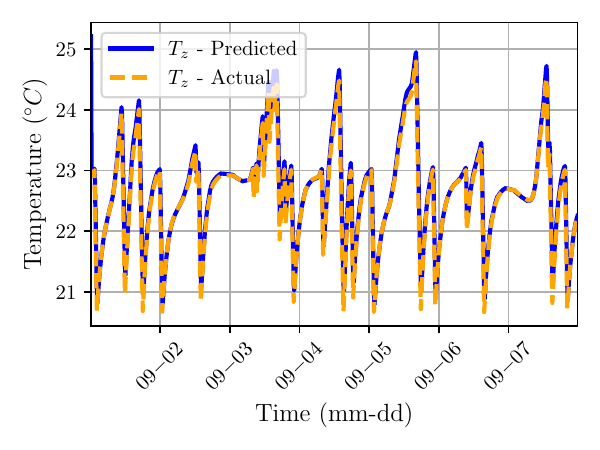}
		\caption{MLE Sim1 R-4.}
		\label{fig:MLE_Sim1_RES4}
	\end{subfigure}
	\begin{subfigure}[t]{0.16\textwidth}
		\centering
		\includegraphics[width=\textwidth]{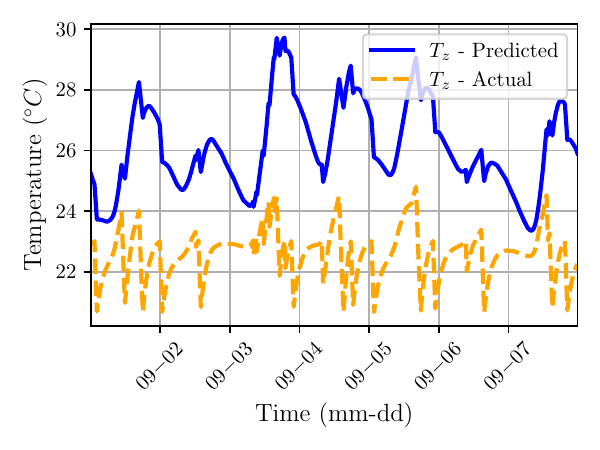}
		\caption{MLE Sim2 R-4.}
		\label{fig:MLE_Sim2_RES4}
	\end{subfigure}
	\begin{subfigure}[t]{0.16\textwidth}
		\centering
		\includegraphics[width=\textwidth]{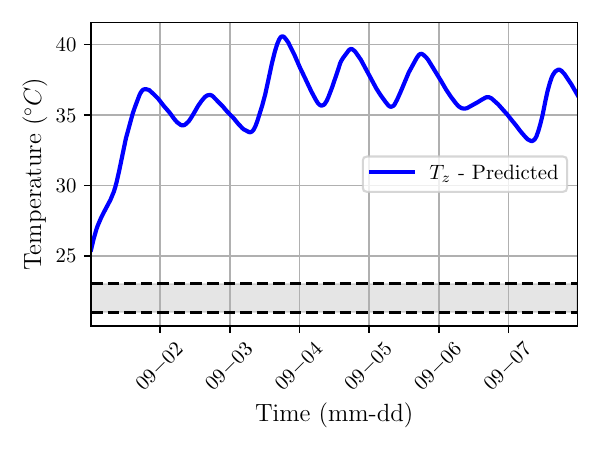}
		\caption{MLE Sim3 R-4.}
		\label{fig:MLE_Sim3_RES4}
	\end{subfigure}
	\begin{subfigure}[t]{0.16\textwidth}
		\centering
		\includegraphics[width=\textwidth]{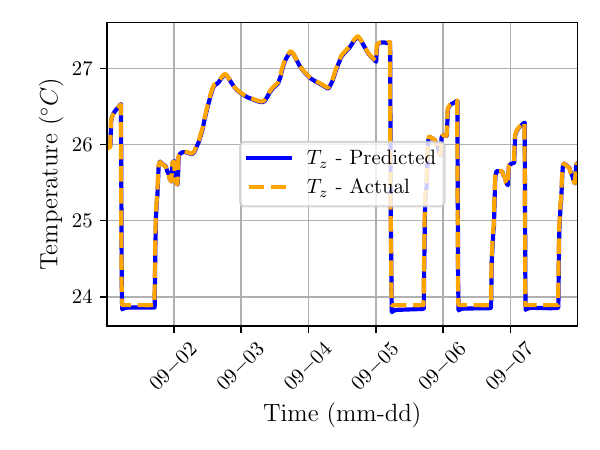}
		\caption{MLE Sim1 C-4.}
		\label{fig:MLE_Sim1_COM4}
	\end{subfigure}
	\begin{subfigure}[t]{0.16\textwidth}
		\centering
		\includegraphics[width=\textwidth]{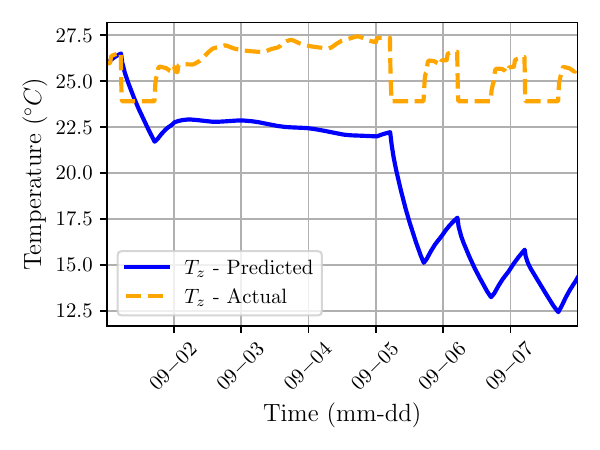}
		\caption{MLE Sim2 C-4.}
		\label{fig:MLE_Sim2_COM4}
	\end{subfigure}
	\begin{subfigure}[t]{0.16\textwidth}
		\centering
		\includegraphics[width=\textwidth]{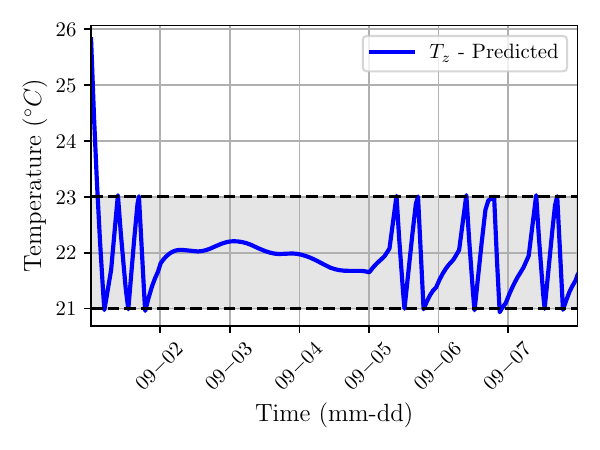}
		\caption{MLE Sim3 C-4.}
		\label{fig:MLE_Sim3_COM4}
	\end{subfigure}

	\begin{subfigure}[t]{0.16\textwidth}
		\centering
		\includegraphics[width=\textwidth]{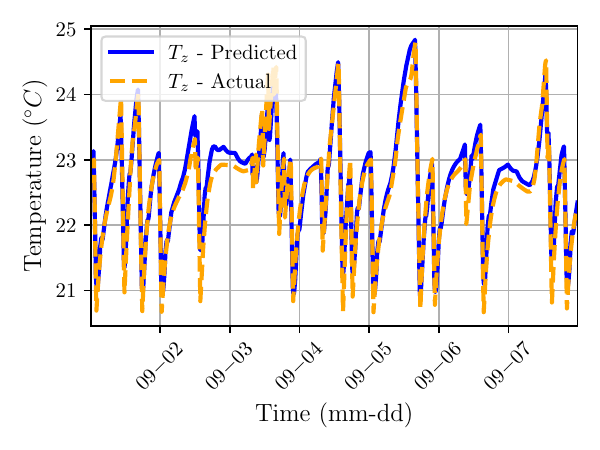}
		\caption{ALS Sim1 R-A.}
		\label{fig:Almon_Sim1_RESA}
	\end{subfigure}
	\begin{subfigure}[t]{0.16\textwidth}
		\centering
		\includegraphics[width=\textwidth]{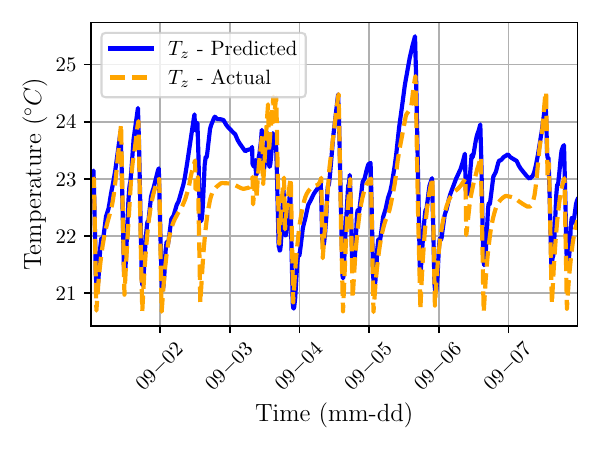}
		\caption{ALS Sim2 R-A.}
		\label{fig:Almon_Sim2_RESA}
	\end{subfigure}
	\begin{subfigure}[t]{0.16\textwidth}
		\centering
		\includegraphics[width=\textwidth]{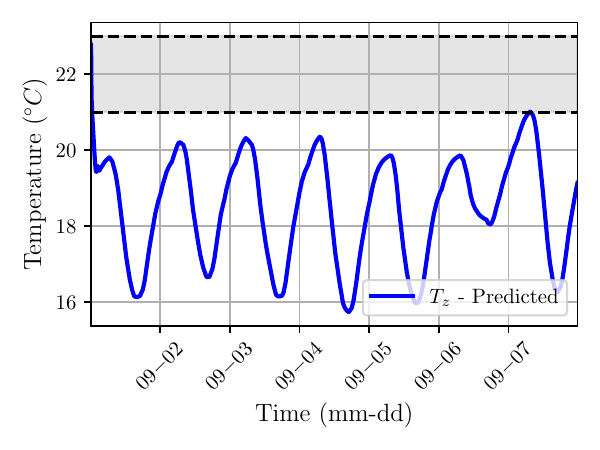}
		\caption{ALS Sim3 R-A.}
		\label{fig:Almon_Sim3_RESA}
	\end{subfigure}
	\begin{subfigure}[t]{0.16\textwidth}
		\centering
		\includegraphics[width=\textwidth]{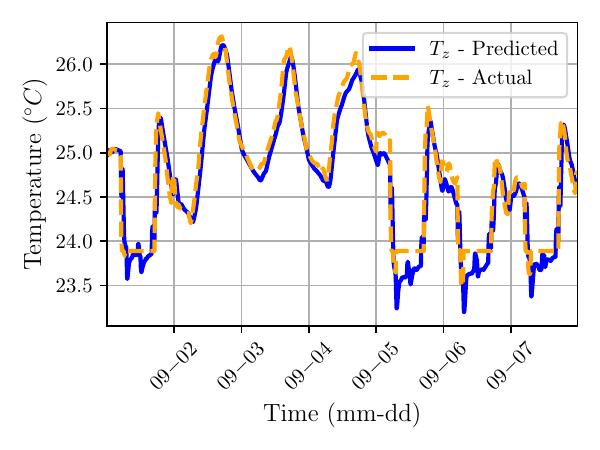}
		\caption{ALS Sim1 C-A.}
		\label{fig:Almon_Sim1_COMA}
	\end{subfigure}
	\begin{subfigure}[t]{0.16\textwidth}
		\centering
		\includegraphics[width=\textwidth]{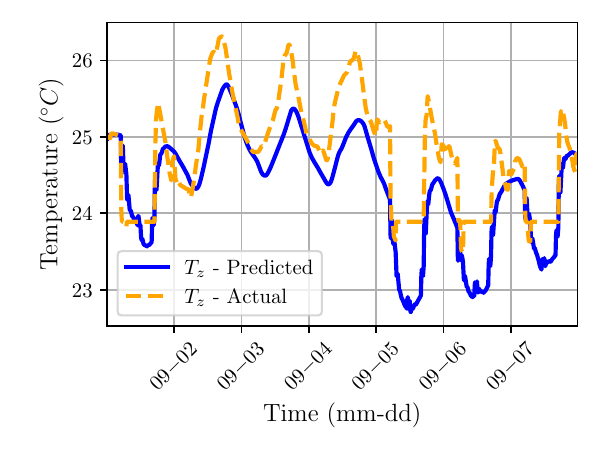}
		\caption{ALS Sim2 C-A.}
		\label{fig:Almon_Sim2_COMA}
	\end{subfigure}
	\begin{subfigure}[t]{0.16\textwidth}
		\centering
		\includegraphics[width=\textwidth]{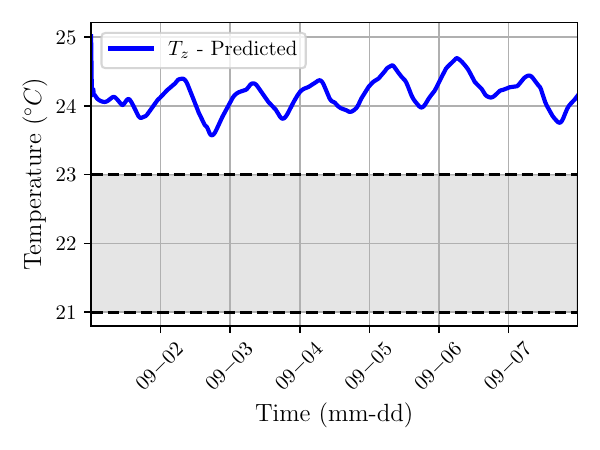}
		\caption{ALS Sim3 C-A.}
		\label{fig:Almon_Sim3_COMA}
	\end{subfigure}
	\caption{Performance comparison for LS, BE, MLE, and ALS methods across house (RES) and building (COM) models for Sim1, Sim2, and Sim3.}
	\vspace{-0.3cm}
    \label{fig:ResultsComparison}
\end{figure*}

\section{Case Study}\label{sec:CaseStudy}
\subsection{Setup}\label{subsec:Setup}

\subsubsection{Data}\label{subsubsec:Data}
House data for model estimation is generated at a 10-minute resolution from a 4-state linear ODE simulation of a single-family detached house \cite{cui2019hybrid}, while commercial building data is generated at a 5-minute resolution using an EnergyPlus~\cite{EnergyPlus} simulation of a prototypical 5-zone small office developed by Pacific Northwest National Laboratory~\cite{PNNLBuildingDataBase}. Both datasets span June 1 to September 30. Both datasets are generated without any artificial excitation under an appropriate control policy, which limits their richness and makes them resemble real-world datasets.

\subsubsection{Training and Testing Data}\label{subsubsec:TrainTestData}
The thermal RC-network estimation methods use 3, 5, 7, 14, and 21 days of data leading up to August 31 for both buildings, while the method for the ALS transformed structured regression model uses data from June 1 through August 31 for estimating the model parameters. Testing data comprises 7 days, from Sept 1-7, for both estimation methods and building models.

\subsubsection{RC-Network Model Architectures}\label{subsubsec:RCNetArchitectures}
For the house, we compare three models: 1. R-1 (1-State, 5-Parameters), 2. R-2 (2-State, 7-Parameters), and 3. R-4 (4-State, 12-Parameters). For the commercial building, we compare two models: 1. C-1 (1-State, 3-Parameters) and 2. C-2 (2-State, 6-Parameters). The models follow (\ref{eq:GB1}) -(\ref{eq:GB2}).

\subsubsection{Structured Regression Model Architectures}\label{subsubsec:AlmonModelArchitectures}
ALS-transformed thermal dynamics models for the house (R-A) and the commercial building (C-A) are given by ~(\ref{eq:ALS_Transform_H}) and ~(\ref{eq:ALS_Transform_B}), respectively.
\begin{align}      
    &T_{z}(k+1) = \alpha_0 + \text{A}(T_{z},6,14,2) + \text{A}(P_{c},0,11,2)    \label{eq:ALS_Transform_H} \\ &  
     + \text{A}(D_{c},6,17,1) + \text{A}(D_{h},6,17,1) , \nonumber \\
     &T_{z}(k+1) = \alpha_0 + \text{A}(T_{z},6,14,2) + \text{A}(P_{c},0,8,2)    \label{eq:ALS_Transform_B} \\ &+ \text{A}(P_{h},0,8,2)  
     + \text{A}(D_{c},6,17,1) + \text{A}(D_{h},6,17,1) . \nonumber 
\end{align}
In eq.~(\ref{eq:ALS_Transform_H}) the lags of $P_{h}$ are not used as the house is not equipped with heating.

\subsubsection{Computation}\label{subsubsec:Computation}
Nonlinear programs for RC-network estimation methods are modeled in CasADi~\cite{CasADi} and solved using IPOPT~\cite{Ipopt}. The ALS-transformed structured regression model's LLS problem is formulated and solved using TSP 5.1~\cite{TSP}.

\subsection{Results And Discussion}\label{sec:ResultsDiscussion}
\begin{figure}[htbp]
    \centering
    \begin{subfigure}[t]{0.45\columnwidth}
        \centering
        \includegraphics[width=\textwidth]{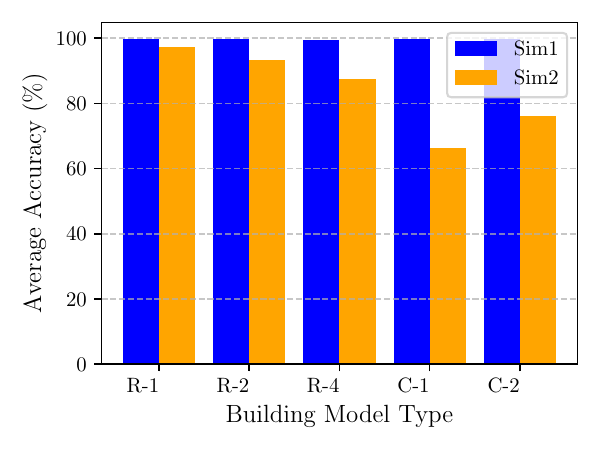}
        \caption{MLE Average Accuracy.}
        \label{fig:MLE_Average_Accuracy}
    \end{subfigure}
    \begin{subfigure}[t]{0.45\columnwidth}
        \centering
        \includegraphics[width=\textwidth]{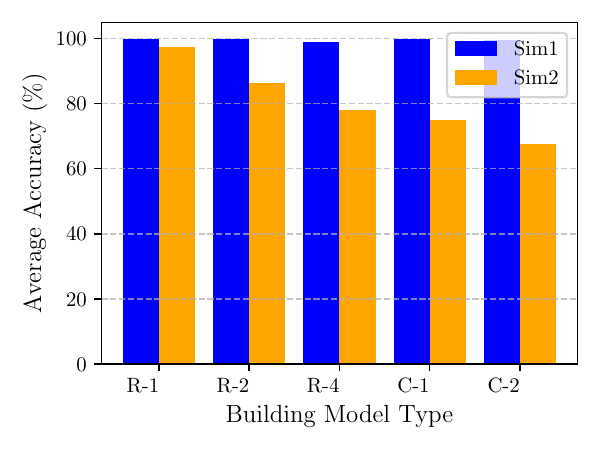}
        \caption{BE Average Accuracy.}
        \label{fig:BE_Average_Accuracy}
    \end{subfigure}
    
    \begin{subfigure}[t]{0.45\columnwidth}
        \centering
        \includegraphics[width=\textwidth]{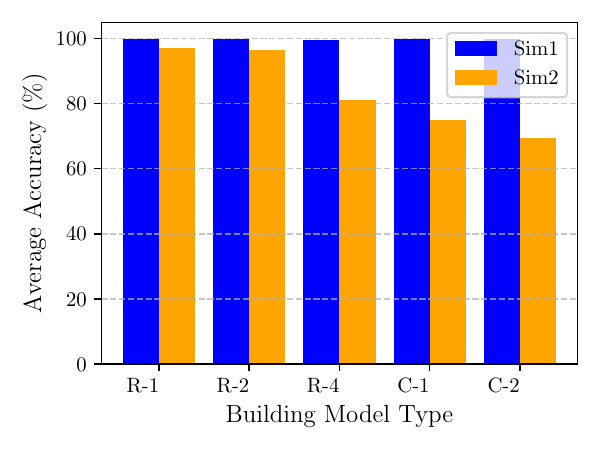}
        \caption{LS Average Accuracy.}
        \label{fig:LS_Average_Accuracy}
    \end{subfigure}
    \begin{subfigure}[t]{0.45\columnwidth}
        \centering
        \includegraphics[width=\textwidth]{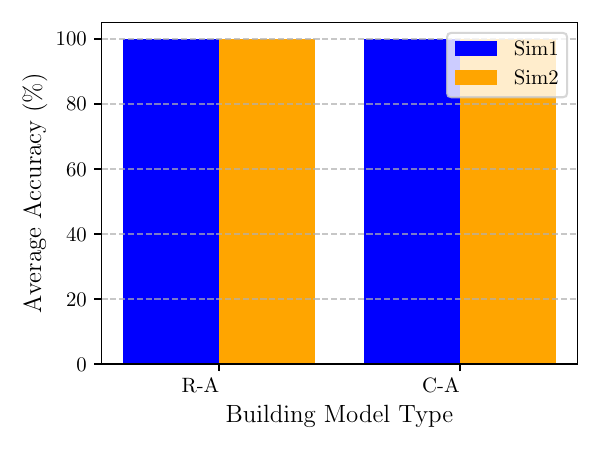}
        \caption{Almon Average Accuracy.}
        \label{fig:ALMON_Average_Accuracy}
    \end{subfigure}

    \caption{Comparison of average accuracy for different estimation methods: MLE, BE, LS, and ALS.}
    \label{fig:Average_Accuracy_Comparison}
\vspace{-0.5cm}
\end{figure}

\subsubsection{Simulation Types and Performance Metric}\label{subsubsec:SimulationTypes}
We evaluate the performance of the estimated parameters ($\ubar{\theta}$) using three system simulation setups based on eq.~(\ref{eq:GM1}): 
1) Simulation Type-1 (Sim1) uses measurable state and input data from the dataset for one-step prediction, 
2) Simulation Type-2 (Sim2) uses only input data from the dataset with complete state feedback for an arbitrary $N$-step prediciton, and 
3) Simulation Type-3 (Sim3) is similar to Sim2 but uses the control input from an arbitrary control policy leading to the most realistic evaluation of the learned dynamics as required in grid-edge applications. 
In this work for Sim3, we implement a basic thermostat control policy with a setpoint of $22^{\circ}\text{C}$ and a deadband of $\pm 1^{\circ}\text{C}$ which is represented by a grey-band in Fig.~\ref{fig:ResultsComparison}.

We use average accuracy, calculated as $100 - \text{MAPE}$ (Mean Absolute Percentage Error), as the performance metric to compare Sim1 and Sim2 across all estimation methods and building types. However, this metric cannot be applied to Sim3, as reference output data under an arbitrary policy is not available for MAPE calculation.

\subsubsection{Performance of Estimation Methods for RC-Network Models}\label{subsubsec:PerformanceGB}
The first three rows of Fig.~\ref{fig:ResultsComparison} illustrates the performance of LS, BE, and MLE methods for RC-Network models across three simulation types. Performance is evaluated on R-4 for house models and C-4 for commercial models, as these performed better in Sim3 compared to other models, which are omitted due to space constraints. The results are based on a testing dataset with models trained on 7 days of data. Larger training datasets showed no significant improvement in results, while the optimization problems became computationally expensive and often failed to converge.

Across LS, BE, and MLE, Sim1 achieves the best performance for both R-4 and C-4 models, as it is the simplest simulation type with no feedback. Sim2 shows the lowest accuracy across all methods and models, as the control data relies on the original thermal state rather than the simulated state fed back during simulation. Sim3 provides the most realistic evaluation of learned dynamics, with only MLE for the C-4 model following the  test control policy, see Fig.~\ref{fig:MLE_Sim3_COM4}. This highlights the challenge of accurately learning system dynamics with these methods, largely due to their inability to handle large datasets effectively which hinders their ability to generalize to arbitrary control policies.

\subsubsection{Performance of ALS Method}\label{subsubsec:PerformanceBB}
The last row of Fig.~\ref{fig:ResultsComparison} illustrates the results based on the testing dataset for the performance of the ALS method for both building models, R-A and C-A. For ALS, both Sim1 and Sim2 perform exceptionally well for both the building models, illustrating the method's ability to learn the thermal dynamics specific to the control policy under which the training and testing data were generated. However, in Sim3, with a different control policy, the ALS method fails to follow this policy for both building models. This limitation can be attributed to the lack of a physics-based structure in the regression models, leading to an inability to generalize to arbitrary control policies.

\subsubsection{Comparison of Structured Regression Models with RC-Network Models}\label{subsubsec:ComparisonBBGB}
The ALS method along with the three RC-Network estimation methods demonstrate similar strong performance in Sim1 for both the building model types. In Sim2, ALS outperforms all other methods for both building models due to its ability to ingest large amounts of data and effectively learn the underlying control policy. Unlike ALS, which leverages LLS for parameter estimation, RC-Network estimation methods use NLS-based approaches that become increasingly challenging to solve optimally as data size and problem complexity grow. For Sim3, all methods show poor performance in general. ALS struggles due to its lack of a physics-based structure, while RC-Network methods fail to utilize large datasets effectively to learn the true dynamics required for arbitrary control policies. Fig.~\ref{fig:Average_Accuracy_Comparison} highlights the average accuracy for Sim1 and Sim2 across different methods and building models. However, this metric is insufficient for assessing a model's performance under Sim3, emphasizing the need for a principled evaluation approach to enable the effective selection of these models for grid-edge applications.


\section{Conclusion}\label{sec:Conclusion}
In this study, we compared RC-Network and regression-based models for building thermal dynamics, along with their estimation methods, across three simulation types. Simulation Type-3 consistently exhibited poor performance across all models and methods, underscoring the limitations of the examined grey-box approaches in capturing generalized dynamics capable of operating under arbitrary control policies, a critical requirement for grid-edge applications. Key challenges include the absence of physics-informed structures in regression-based models and scalability issues in RC-Network estimation methods as data volume increases. Future work will focus on overcoming these challenges by leveraging advancements in Scientific Machine Learning to develop more scalable, physics-informed models and defining a new performance metric for evaluating the learning of generalized building dynamics.


\section*{Acknowledgment}
This work is funded by the NSF award 2208783.


\bibliographystyle{IEEEtran}
\input{output.bbl}


\end{document}